\documentclass[aps,prb,showpacs,twocolumn,reprint,superscriptaddress]{revtex4-1}
\usepackage{amsmath}
\usepackage{amssymb}
\usepackage{graphicx}
\usepackage{color}
\usepackage{multirow}
\usepackage{dcolumn}
\usepackage{textcomp}
\begin{document}
\title{Layer $k$-projection and unfolding electronic bands at interfaces}
\author {Mingxing Chen}
\affiliation{School of Physics and Electronics, Hunan Normal University, Changsha, Hunan 410081, China}
\author {M. Weinert}
\affiliation{Department of Physics, University of Wisconsin, Milwaukee, Wisconsin 53211, USA}

\date{\today}

\begin{abstract}
The $k$-projection method provides an approach to separate the contributions from different constituents
in heterostructure systems, and can act as an aid to connect the results of experiments and calculations. 
We show that the technique can be used to ``unfold'' the calculated electronic bands of interfaces and supercells, and
provide local band structure by integrating the projected states over specified regions of space, a step that
can be implemented efficiently using fast Fourier transforms.  
We apply the
method to investigate the effects of interfaces in heterostructures consisting of a graphene bilayer on H-saturated
SiC(0001), BAs monolayer on the ferromagnetic semiconductor CrI$_3$, silicene on Ag(111), and to the
Bi$_2$Se$_3$ surface.  Our results
reveal that the band structure of the graphene bilayer around the Dirac point is strongly dependent on the
termination of SiC(0001): on the C-face, the graphene is $n$-doped and a gap of $\sim$0.13 eV is opened, whereas
on the Si-face, the graphene is essential unchanged and neutral. We show that for BAs/CrI$_3$, the magnetic proximity effect can effectively induce a spin splitting up to about
50 meV in BAs.  For silicene/Ag(111), our calculations reproduce the angle-resolved photoemission
spectroscopy results, including linearly dispersing bands at the edge of the
first Brillouin zone of Ag(111); although these states result from the interaction between the silicene
overlayer and the substrate, we demonstrate that they are not Dirac states.
\end{abstract}

\pacs{71.20.-b,73.20.-r,73.22.Pr}

\keywords{ab initio calculations, Band structure; band unfolding, interfaces}

\maketitle
\section{INTRODUCTION}
Doping has been an important means of tailoring electronic properties of materials, and exploring novel
physical phenomena, such as doping Mott insulators to obtain high T$_\mathrm{c}$
superconductors,\cite{bednorz_1986} doping Ge by Sn to obtain direct semiconductors, \cite{Gupta2013} and
realizing quantum anomalous Hall effect in topological insulators by doping magnetic
impurities.\cite{Yu2010,Chang2013} As the thickness of materials approach the atomic limit, the interface
between the material and the substrate plays a critical role in determining its atomic structure and electronic
properties.  The increased interest in interface effects has motivated in part by experimental realization of
stable monolayer systems such as graphene and transition-metal dichalcogenides (TMDs).  For example, the
magnetic proximity effect has been used to manipulate spin polarization, spin-valley polarization, and explore
quantum anomalous Hall effect in the monolayers by placing them onto the surface of magnetic
semiconductors.\cite{Yang2013,Wei2016,Li2015Giant,Zhang2016Large,Qiao2014,Wang2015,Zhang2015Quantum} The
research has been further expanded by the development of van der Waals (vdW) heterostructures, enabling
the design of materials with properties distinct from their constituents.\cite{Novoselov20162D} 

First-principles calculations have played an important role in understanding the effects of doping and
interfaces on the electronic structures of materials.  There are a number of approaches that are
typically used to model doping in solids: (\textit{i}) The virtual crystal approximation
(VCA)\cite{nordheim} and the coherent potential approximation (CPA)\cite{Gyorffy1972Coherent} methods in
which an ``averaged'' problem is solved; (\textit{ii}) supercell approaches where defects are
periodically repeated; and (\textit{iii}) impurity Green's function (GF) methods
\cite{KKR-defects,KKR-full} that treat an isolated defect (or cluster) embedded in a host. 
Similarly, at heterostructure interfaces, lattice mismatch
between the two constituents and/or interface reconstructions can lead to interface structures that are
intrinsically supercells relative to the bulk systems.  
In both the
supercell and Green's function methods the translational symmetry is reduced (lost in the case of the
GF impurity calculations) from that of the bulk, and leads to ``band folding'', including of the bulk
bands. Given a set of calculations, to separate the effects of the dopants, interfaces, etc., requires
``unfolding'' the bands, i.e., determining the correspondence between the wave functions of the bulk and
defect/interface systems.

\begin{figure}
  \includegraphics[width=1.0\columnwidth]{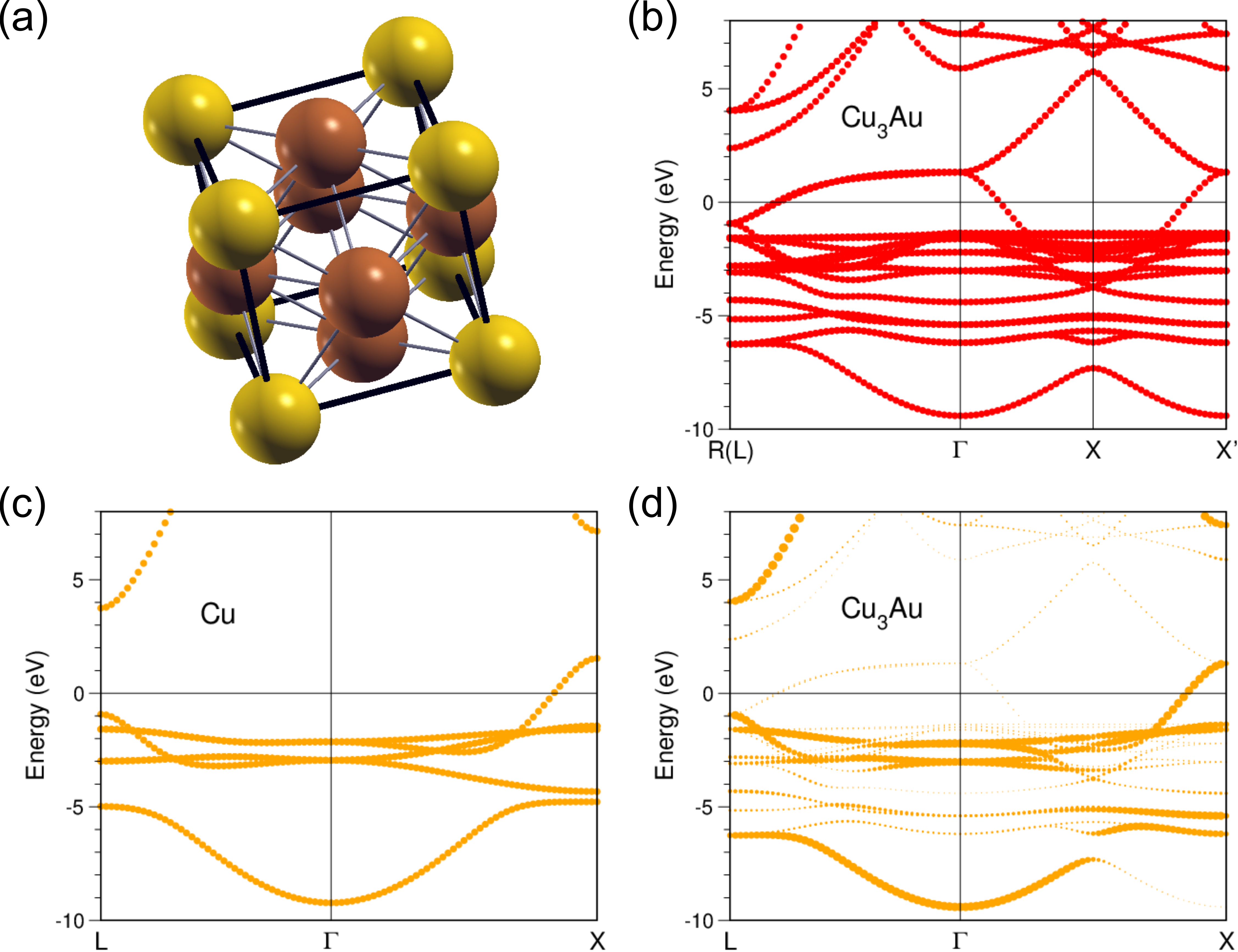}
  \caption{
  (a) Simple cubic Cu$_3$Au structure.
  (b) Calculated band structure of Cu$_3$Au along high symmetry directions, 
  (c) Band structure for fcc Cu.
  (d) Cu$_3$Au $k$-projected (unfolded) bands. The sizes of the filled circles are the
$k$-projected weights of the states shown in (a).
The Fermi level is set to zero.
  }
 \label{fig1}
\end{figure}

This correspondence is a general issue and can provide insights into the the underlying physics. An early
application\cite{Weinert1988} of band unfolding was motivated by the observation that
the photoemission of simple cubic
Cu$_3$Au closely resembles that of fcc Cu: the calculated bands of Cu$_3$Au,
Fig.~\ref{fig1}(b), show significantly more bands with dispersions different that of
Cu, Fig.~\ref{fig1}(c). However, applying a $k$-projection technique to the Cu$_3$Au bands --- treating
the Cu$_3$Au structure as an ordered supercell impurity phase, Fig.~\ref{fig1} --- not only
recovers the fcc-like band structure, but also reveals which ``Cu'' states hybridize and their relative
weights, going beyond simply ``unfolding'' the bulk bands. Thus, $k$-projection 
can provide key insights to the underlying physics.

A number of strategies have been developed to unfold the electronic bands from supercell calculations,
often based on plane-wave methods or tight-binding
methods.\cite{Weinert1988,Qi2010Epitaxial,Boykin2005,Ku2010,Lee2013,huang2014,Roser2014,APL_step}
In principle, the procedure is straightforward, especially for plane wave-based methods, but can become
time-consuming if done naively, especially in cases such as separating overlayer and substrate contributions.
In this paper we present details of an efficient layer $k$-projection method that
allows us to study the local band structure,\cite{Qi2010Epitaxial,Chen2014} and the relationship to 
angle-resolved photoemission spectroscopy (ARPES)\cite{FeSe_STO,LiOHFeSe}
and scanning tunneling microscopy/spectroscopy (STM/STS)\cite{Zhang2018} experiments.

The paper is organized as follows.  In Section II, we discuss the $k$-projection band unfolding technique and
computational details of density-functional theory (DFT) calculations.  In Section III, we present
applications of the method to four interface structures: graphene bilayer (gr-2L) on H-saturated SiC(0001),
BAs monolayers on the ferromagnetic semiconductor CrI$_3$ monolayer, silicene on Ag(111), and the
bulk-surface decomposition of the Dirac state on the topological insulator Bi$_2$Se$_3$. 

\begin{figure}
\includegraphics[width=0.95\columnwidth]{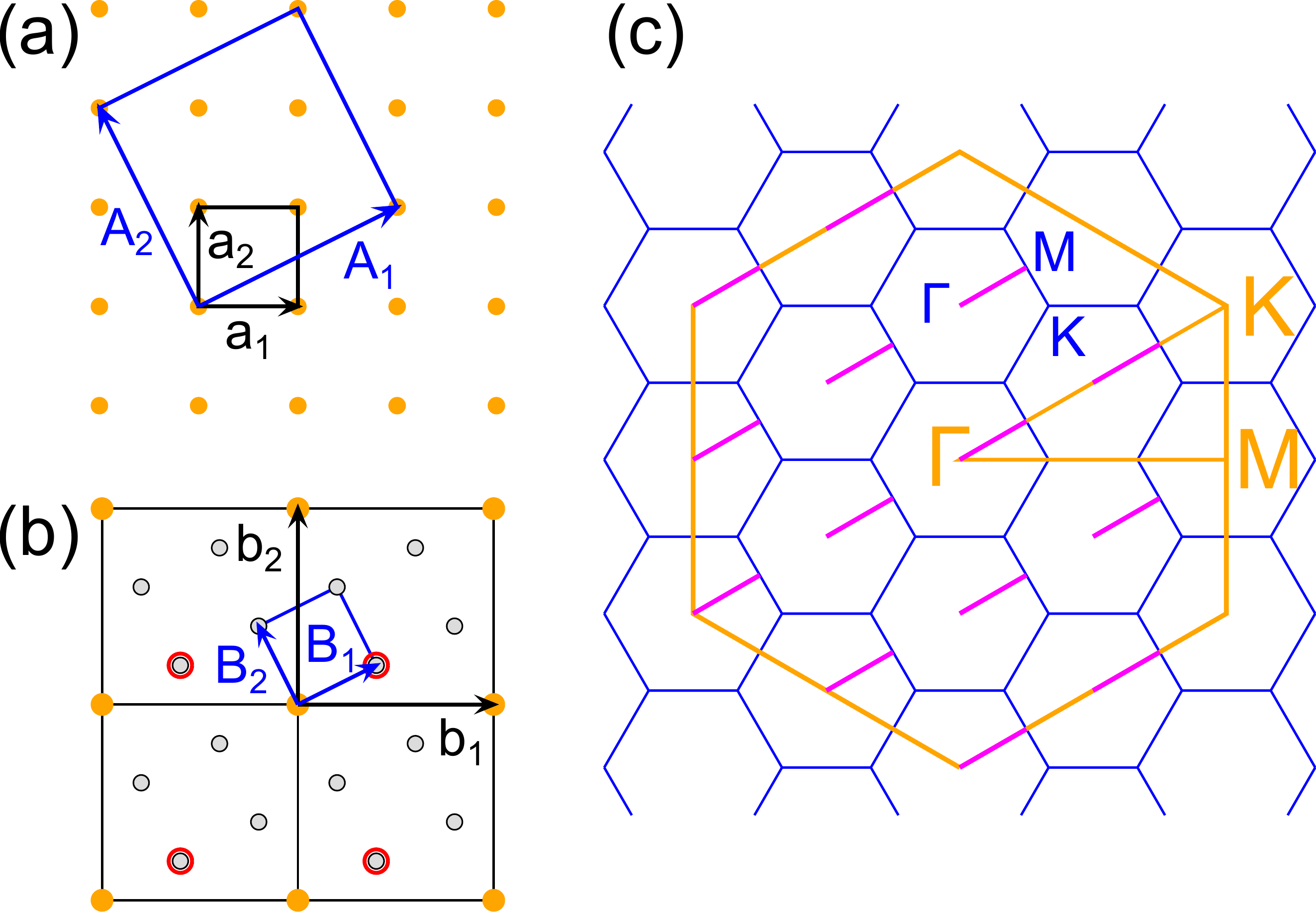}
\caption{(a) Direct lattice, with $\mathbf{a}_i$ ($\mathbf{A}_i$) the basis vectors of the primitive
(supercell).
(b) The corresponding reciprocal lattice basis vectors. The orange circles are the reciprocal lattice vectors
$\mathbf{G}$ of the primitive lattice; the red marked points are reciprocal lattice vectors of the supercell
given by $\mathbf{G_s} = \mathbf{G} + \mathbf{B}_1$, corresponding to $\mathbf{k}=\mathbf{B}_1$ of the
primitive cell.
(c) Relationship between the 1$\times$1 (orange) and $2\sqrt{3}$$\times$$2\sqrt{3}$ (blue) hexagonal Brillouin zones.
Calculations along $\Gamma$--M (magenta) for the supercell correspond to 12 different lines in the BZ of the
1$\times$1 structure.
}
\label{fig2}
\end{figure}

\section{METHOD AND COMPUTATIONAL DETAILS}

In a system with translational symmetry defined by the set (of order $h$) translation operators
$\hat{T}_\mathbf{t}$, the irreducible representations (irreps) can be labeled by $\mathbf{k}$ within the
first Brillouin zone.  (The set of translations $\lbrace\mathbf{t}\rbrace$ are integral multiples
of the direct lattice
vectors $\mathbf{a}_i$, while reciprocal lattice vectors $\mathbf{G}$ as well as $\mathbf{k}$ are given in
terms of $\mathbf{b}_j$.) The character of
the $\hat{T}_\mathbf{t}$ for the $\mathbf{k}$ irrep is
$\chi_\mathbf{k}(\mathbf{t})= e^{i\mathbf{k}\cdot\mathbf{t}}$. Applied to a function $\psi$, the projection operator
\begin{equation}
 \hat{P}_\mathbf{k} = \frac{1}{h}\, \sum_\mathbf{t}\, \chi^*_\mathbf{k}(\mathbf{t})\, \hat{T}_\mathbf{t},
\end{equation}
will project out that part $\psi_\mathbf{k}$ that transforms as the $\mathbf{k}$th irrep
\begin{equation}
 \psi_\mathbf{k} = \hat{P}_\mathbf{k}\, \psi = \frac{1}{h} \,
\sum_\mathbf{t} \chi^*_\mathbf{k}(\mathbf{t}) \hat{T}_\mathbf{t} \psi ,
\end{equation}
and $\hat{P}_\mathbf{k} \psi_\mathbf{k} = \psi_\mathbf{k}$. Since $\sum_\mathbf{k} \hat{P}_\mathbf{k} = 1$,
any function can be decomposed into pieces that transform as the irreps labeled by $\mathbf{k}$
\begin{equation}
\psi = \sum_\mathbf{k} \psi_\mathbf{k} ,
\end{equation}
and $\langle\psi_{\mathbf{k}}|\psi_{\mathbf{k}}\rangle$ measures the relative weight of this $k$ to
$\psi$. 
Note that this is general, and the function $\psi$ itself need not actually possess any translational
symmetry, thus allowing this $k$-projection to be applied even to single impurity (GF) calculations where
the translational symmetry is explicitly lost. Since the irreps labeled by $\mathbf{k}$ and
$\mathbf{k}$+$\mathbf{G}$ are the same, the projection operator puts $\psi_\mathbf{k}$ into the form
\begin{equation}
 \psi_\mathbf{k}(\mathbf{r}) = \sum_\mathbf{G} \psi_\mathbf{k}(\mathbf{G}) e^{i(\mathbf{k}+\mathbf{G})\cdot\mathbf{r}}
,
\end{equation}
where $\psi_\mathbf{k}(\mathbf{G})$ is related to the standard Fourier transform $\psi(\mathbf{q})$,
$\mathbf{q}=\mathbf{k}+\mathbf{G}$,
\begin{equation}
 \psi(\mathbf{q}) = \frac{1}{\Omega} \,
     \int \, d\mathbf{r} \, \psi(\mathbf{r}) e^{-i\mathbf{q}\cdot\mathbf{r}} .
\label{eq-fourier}
\end{equation}
This connection to the Fourier coefficients $\psi(\mathbf{q})$ provides a prescription for making the
$k$-projected decomposition. When the system possess, at least approximately, the translational symmetry
defined by the $\mathbf{a}_i$, the Fourier coefficients will have peaks at the corresponding $\mathbf{q}$
values; when this is not the case, i.e., the projection is done on the ``wrong'' translational symmetry,
the weights will be spread throughout $\mathbf{q}$-space.

In practice, calculations are often done in a supercell geometry relative to a primitive cell (c.f.,
Fig.~\ref{fig2}) with
direct lattice vectors $\mathbf{A}_i = \sum_j n_{ij} \mathbf{a}_j$ and reciprocal lattice basis vectors
$\mathbf{B}_j$, $\mathbf{b}_i = \sum_j m_{ij} \mathbf{B}_j$, where
$n_{ij}, m_{ij}$ are integers. In this case, because the translational symmetries are commensurate, the
Fourier coefficients, Eq.~(\ref{eq-fourier}), are non-zero at only specific $\mathbf{q}$ values.
A state calculated at $\mathbf{k_s}$ will (possibly) have non-zero Fourier coefficients for $\mathbf{k_s}+\mathbf{G_s}$,
but $\mathbf{G_s}$ may not be a reciprocal lattice vector of the primitive cell:
\begin{eqnarray*}
  \mathbf{G_s} & = & \sum_i M_i \mathbf{B_i} \\
               & = & \sum_j \left( \sum_i M_i (\mathbf{B}_i\cdot\mathbf{a}_j)\right) \\
               & = & \sum_j (m_j + \kappa_j) \mathbf{b}_j \\
               & = & \mathbf{G} + \mathbf{\kappa}  \equiv \mathbf{G} + \mathbf{G_s}^0 ,
\end{eqnarray*}
where $M_i$, $m_j$ are integers, $\mathbf{a}_i\cdot\mathbf{b}_j=\delta_{ij}$, $\mathbf{G}$ belongs to the
primitive lattice, and
$\mathbf{\kappa} = \sum_j \kappa_j \mathbf{b}_j = \mathbf{G_s}^0$ is in the first BZ of the primitive cell
($\kappa_j$ is fractional), but is equal to a
reciprocal lattice vector of the supercell. If the supercell is $N$ times as
large as the primitive, then there will be $N$ distinct $\mathbf{G_s}^0$, and hence a function calculated at
$\mathbf{k_s}$ in the supercell can be decomposed ($k$-projected) into pieces corresponding to $\mathbf{k} =
\mathbf{k_s} + \mathbf{G_s}^0$ of the primitive cell.

For an ideal supercell, this decomposition is exact since it is a simple consequence
of translational symmetry and recovers the primitive band structure with
$|\langle\psi_{\mathbf{k}_s}|\psi_{\mathbf{k}}\rangle|=1$ or 0; 
for defect systems and interfaces, the norm will be between zero and one.  

The above scheme requires determining $\psi_\mathbf{k}$ and then the weight
$\langle\psi_\mathbf{k}|\psi_\mathbf{k}\rangle$ for $\mathbf{k}$ in the ``primitive'' cell. For bulk
defect systems, the spatial integration is over the whole space. For interface systems, or for modeling STM
or other probes that measure local properties, the integration volume entering the weight calculation should
be restricted.

Our implementation, which uses FFTs, provides an efficient approach and has been used in plane wave
(pseudopotential and PAW) methods and augmented methods (FLAPW and LASTO), and also can be adapted to LCAO ones.
The first step is to determine $\psi_\mathbf{k}$ from the wave function calculated in the supercell at $\mathbf{k_s}$,
$\psi_\mathbf{k_s}$. For plane wave-based methods, all the $\mathbf{G_s}$ wave function coefficients not
corresponding to $\mathbf{k}$ and $\mathbf{G}$ are simply zeroed out; for other basis sets (e.g., LCAO) or
when the lattices are not commensurate, an initial step is to evaluate the wave function in real space on the
FFT mesh, back transform to reciprocal space, and then zero out the appropriate coefficients. 

The projected wave function $\psi_\mathbf{k}$ at this point can be put back into the normal basis used to
represent the wave function, but now with modified coefficients. Then the standard machinery used to calculate
overlaps (e.g., local density of states) can be used to calculate the weight, restricting the integration to a
particular region of space as necessary. This restriction corresponds to including a step function
$U(\mathbf{r})$ (non-zero only in the region of interest) in the calculation of the weight. For plane wave
components (including augmented methods like the FLAPW), this convolution is easily done: (\textit{i}) FFT
$\psi_\mathbf{k}$ to real space; (\textit{ii}) square, $\rho_\mathbf{k}(\mathbf{r}) =
|\psi_\mathbf{k}(\mathbf{r})|^2|$; (\textit{iii}) back transform to reciprocal space to obtain the Fourier
coefficients $\rho_\mathbf{k}(\mathbf{G})$; and finally (\textit{iv}) performing the sum $\sum_\mathbf{G}
U^*(\mathbf{G}) \rho_\mathbf{k}(\mathbf{G})$. Note that the maximum $|\mathbf{G}|$ in the sum is simply
twice the plane wave cutoff of the wave functions, and hence the sum is exact despite the fact that the step
function has a slow ($G^{-1}$) convergence. For layer regions of space, $z_1<z<z_2$ (with
$\mathbf{a}_3=a_3 \hat{\mathbf{z}}$),
the Fourier coefficients of the
step function are particularly simple
\begin{eqnarray*}
  U(\mathbf{G})& = &\delta_{G_\|,0} \, \frac{1}{a_3} \int_{z_1}^{z_2} dz \, e^{-ig_z z} \\
               & = &\delta_{G_\|,0} \, \frac{2e^{-ig_z(z_2+z_1)/2}}{g_za_3}
        \sin \frac{g_z(z_2-z_1)}{2} .
\end{eqnarray*}
Our implementation is different from Ref.~\onlinecite{APL_step} that directly uses a projector built from the Heaviside function ($\Theta(z)$) 
to seperate contributions of different layers.

\begin{figure*}
  \includegraphics[width=0.90\textwidth]{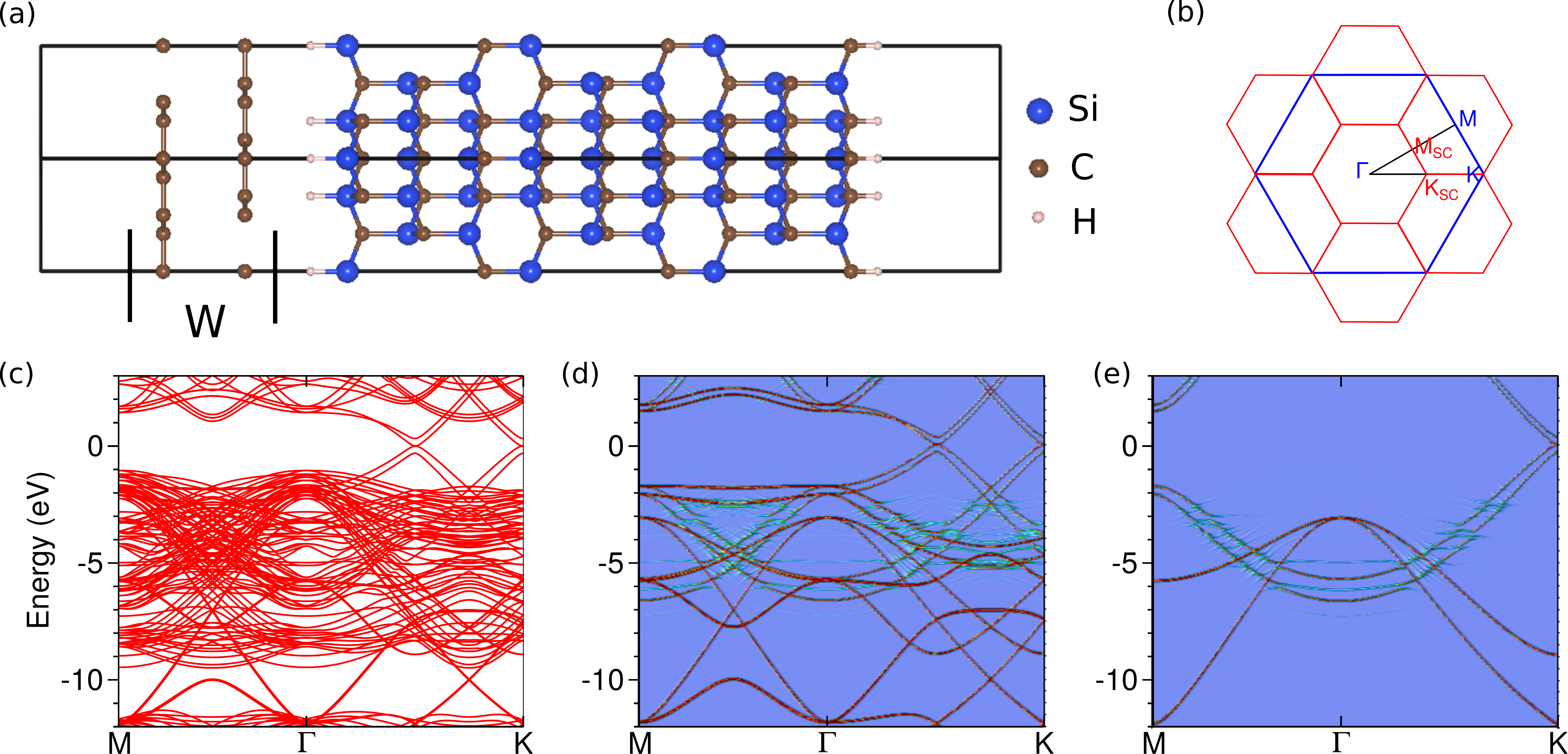}
  \caption{Unfolded band structures for gr-2L/SiC(0001). 
  (a) Geometry of gr-2L on Si-terminated SiC(0001). 
  The gray, brown and blue balls denote H, C, and Si atoms, respectively.
  The graphene bilayer are in AB stacking with an interlayer distance of 3.35 \AA.
  W represents the spatial window for the band unfolding for the gr-2L.
  (b) High symmetry points in the BZs of the supercell (red) and the primitive cell (blue).
  (c) Electronic bands calculated in the supercell along the high symmetry directions of the 1$\times$1 cell.
The boundaries of the 2$\times$2 supercell, M$_\mathrm{SC}$ and K$_\mathrm{SC}$, are indicated by dashed lines.
  (d) Same as (c), but with the bands weighted by the contribution in the graphene layers.
  (e) Bands $k$- and layer-projected to the 1$\times$1 graphene cell.
  The Fermi level is set to zero. 
  }
 \label{fig3}
\end{figure*}

Prior to the band unfolding, we carried out DFT calculations using the Vienna Ab Initio Simulation Package
\cite{kresse1996} for the proposed systems.  The pseudopotentials were constructed by the projector augmented
wave method.\cite{bloechl1994,kresse1999} van der Waals dispersion forces between the adsorbate and the
substrate were accounted for through the optPBE-vdW functional by using the vdW-DF method developed by
Klime\v{s} and Michaelides.\cite{klimes2010,klimes2011} The interface structure is modeled in terms of a
repeated slab, separated from its periodic images by 10 \AA\ vacuum regions.  For gr-2L/SiC(0001), the slab is
composed of a $\sqrt{3}$$\times$$\sqrt{3}$ supercell of H-saturated SiC(0001) and a  2$\times$2 supercell of the
gr-2L.  For BAs/CrI$_3$,  a  2$\times$2 supercell of BAs on a  1$\times$1 unit cell of CrI$_3$ is used, and
for silicene/Ag(111) a 3$\times$3 supercell of silicene on 4$\times$4 Ag(111) is chosen, resulting in a small
lattice mismatch.  To avoid artificial interactions between the polar slabs, two such slabs, oppositely
oriented with mirror symmetry, are placed in each supercell for gr-2L/SiC(0001) and BAs/CrI$_3$,  while for
silicene/Ag(111) the overlayers are symmetrically placed on both sides of the substrate.  To sample the surface
BZs a 12$\times$12 $\Gamma$-centered Monkhorst-Pack $k$-point mesh was used for gr-2L/SiC(0001), 15$\times$15
for BAs/CrI$_3$, and 6$\times$6 for silicene/Ag(111), respectively.  Plane-wave energy cutoffs of 700 eV, 350
eV, 240 eV were used for the electronic structure calculations of the above three interfaces, respectively.
The Bi$_2$Se$_3$ calculations were done using the Full-potential Linearized Augmented Plane Wave \cite{FLAPW}
method for 10 QLs in a single film geometry, i.e., no periodic images, and included spin-orbit. The plane wave
cutoffs were 220 and 2000 eV for the wave functions and density/potential, respectively, and a 12$\times$12
$k$-point mesh was used.

\section{APPLICATIONS TO INTERFACES}

\subsection{Effects of surface termination: graphene bilayer on 6H-SiC}

\begin{figure}
  \includegraphics[width=0.95\columnwidth]{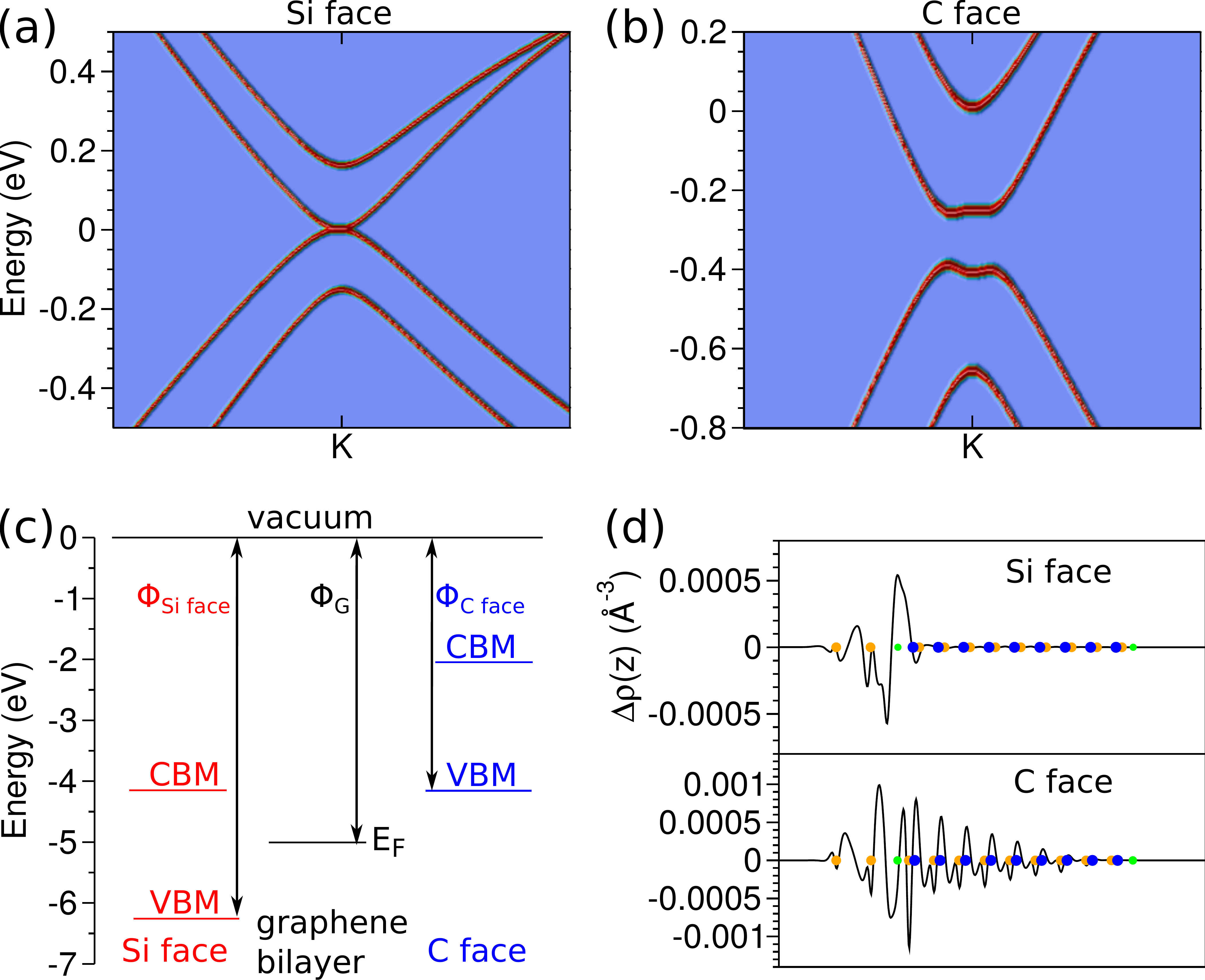}
  \caption{(a), (b) Unfolded band around K for gr-2L on the two faces of SiC(0001).
  (c) Work functions and band alignments of the isolated systems.
  (d) Planar-averaged charge density difference $\Delta \rho(z)$ for gr-2L/SiC(0001).
  Dots show the positions of atoms. Blue, orange, and green dots denote Si, C, and H atoms, respectively. 
  The Fermi level is set to zero.
  }
 \label{fig4}
\end{figure}

\begin{figure*}[ht]
  \includegraphics[width=0.90\textwidth]{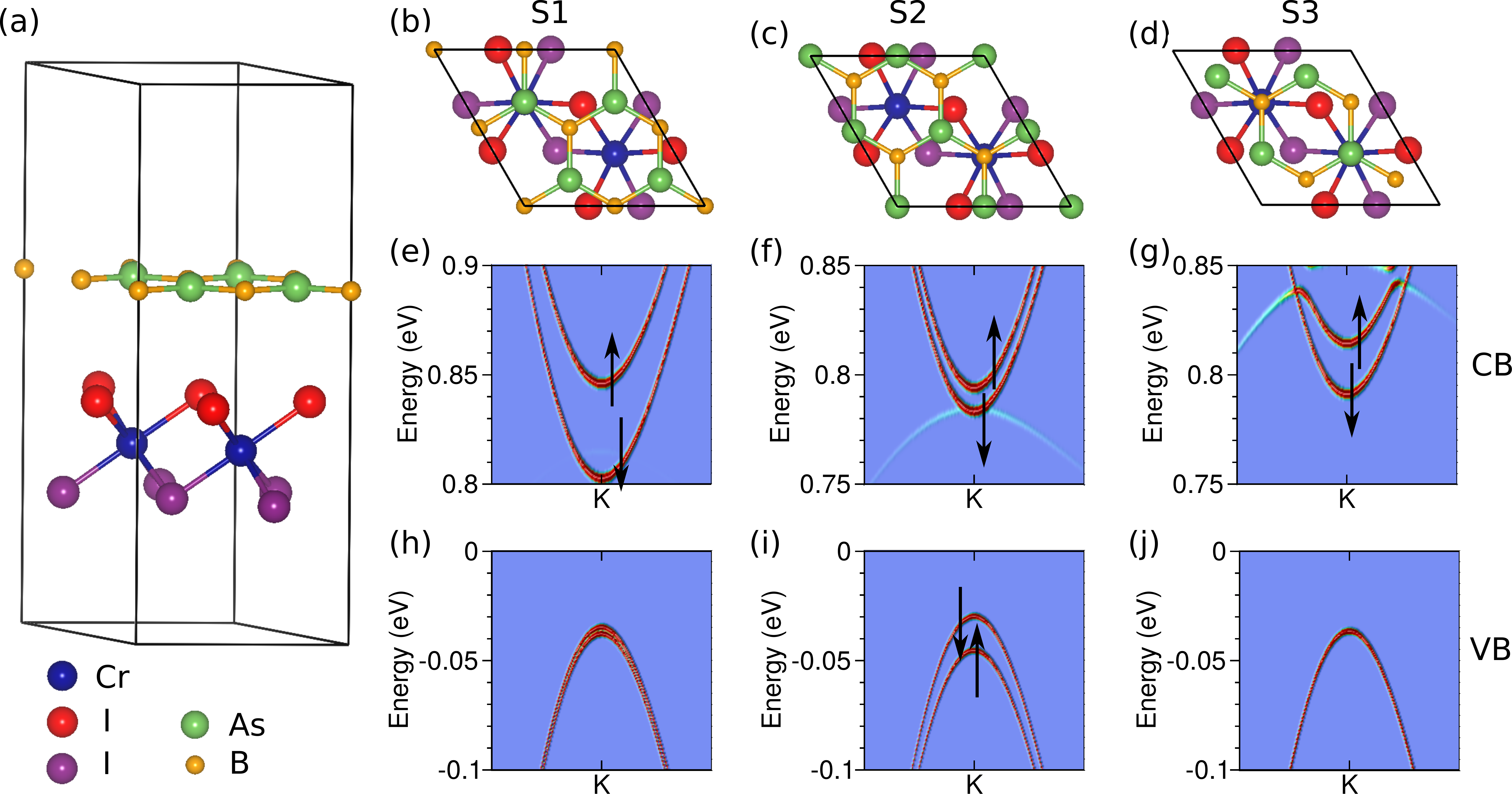}
  \caption{Geometric and electronic structures of BAs/CrI$_3$.
  (a) Side view of (2$\times$2) BAs and CrI$_3$ heterostructure.
  (b)--(d) Top views for three different stackings, which are labeled as S1, S2, and S3, respectively.
  For each configuration, the unfolded (e)-(g) conduction and ((h)-(j) valence band structures were shown
below.
  The Fermi level is set to zero. 
  }
 \label{fig5}
\end{figure*}

The 6H-SiC has two different (0001) surface terminations, either Si or C, and therefore, two different types of
interface structures for bilayer graphene.  Here we consider the situation where the two faces are saturated by
H to model those  experiments where H is used to passivate the interfaces between graphene layers and SiC(0001)
surfaces;\cite{Riedl2009,Riedl2010,Rajput2013} the case of graphene on the Si-face, including the role of
the graphene buffer layer, has been treated earlier.\cite{Qi2010Epitaxial}
Fig.~\ref{fig3}(c) shows the calculated bands for gr-2L on the Si-face along the high symmetry lines extended
into the second BZ of the supercell (c.f., Fig~\ref{fig3}(b)). Because the graphene and substrate bands overlap,
it is difficult to separate the various contributions. To focus on the graphene bands, we first do a layer
projection, integrating over the region defined by W in Fig.~\ref{fig3}(a). These bands, Fig.~\ref{fig3}(d),
while having removed the substrate, still do not simply resemble the free bilayer graphene. Along $\Gamma$-M,
the simple folding about M$_\mathrm{SC}$ clearly visible, but this is not the case along $\Gamma$-K, which
corresponds to two high symmetry lines ($\Gamma$--K$_\mathrm{SC}$ and K$_\mathrm{SC}$--K$_\mathrm{SC}'$) of
the 2$\times$2 supercell calculation. Finally doing the $k$-projection to the 1$\times$1 cell,
Fig.~\ref{fig3}(e), recovers bands that closely resemble the free-standing bilayer ones. Analysis of these
bands provide insight into the interactions of the graphene and the substrate; for example, the mini-gaps
are the result of the hybridization between the graphene and substrate.

The $k$-projected (unfolded) bands around the K point for gr-2L on the Si- and C-faces are shown in
Figs.~\ref{fig4}(a) and (b), respectively. On the H-passivated Si-terminated surface, 
Fig.~\ref{fig4}(a), the interaction with the substrate has only a minor effect on these bands, opening a
gap at K of less than about 10 meV, with the Fermi level located in the middle of the gap.
On the C-face, Fig.~\ref{fig4}(b), the induced gap is $\sim$0.13 eV, an order of magnitude larger than on the
Si-face, and the graphene bilayer is n-doped with the Fermi level located at about 0.3 eV above the gap.

Using the picture proposed for heterostructures composed of silicene (germanene) monolayers and substrates,
the perturbation on the bands of the overlayer about the Dirac point depends on the strength of the hybridization ($V_{int}$) and 
the energy differences ($\Delta E$) between the substrate states and the Dirac point.\cite{chen_design} 
A large $V_{int}$ and a small $\Delta E$ favor a strong perturbation to the Dirac states.
Since gr-2L and the substrate interact mainly via vdW-type bonding, $V_{int}$ is expected to be rather small,
which is supported by the observations that the overall $k$-projected bands do not show large changes that
could be attributed to strong bonding. 
Therefore, the difference in the band structure for gr-2L on different surfaces of SiC 
originates from the difference in $\Delta E$ between the two configurations
determined by the band alignment of the two constituents. 
Fig.~\ref{fig4}(c) depicts the band alignment of the free-standing gr-2L and SiC(0001). 
The Dirac point lies in the gap of the Si-face when their bands align, but
lies below the valence band on the C-face.
This alignment gives rise to a much larger $\Delta E$ when the gr-2L is placed on the Si-face than on the C-face 
($\Delta E$ is expected to be extremely small for the case of the C-face).
Thus, the Fermi level can be expected to cross the Dirac point of the gr-2L on the Si-face and
the states of the gr-2L near the Dirac point experience smaller perturbation.
Figure~\ref{fig4}(d) shows the planar-averaged charge density difference for gr-2L on the two different faces:
The charge polarization in the graphene is much larger on the C-face than on the Si-face,
indicating a stronger dipole field between the gr-2L and the C-face than the Si-face.

\subsection{Magnetic proximity effect in BAs/CrI$_3$}

\begin{figure*}
  \includegraphics[width=0.90\textwidth]{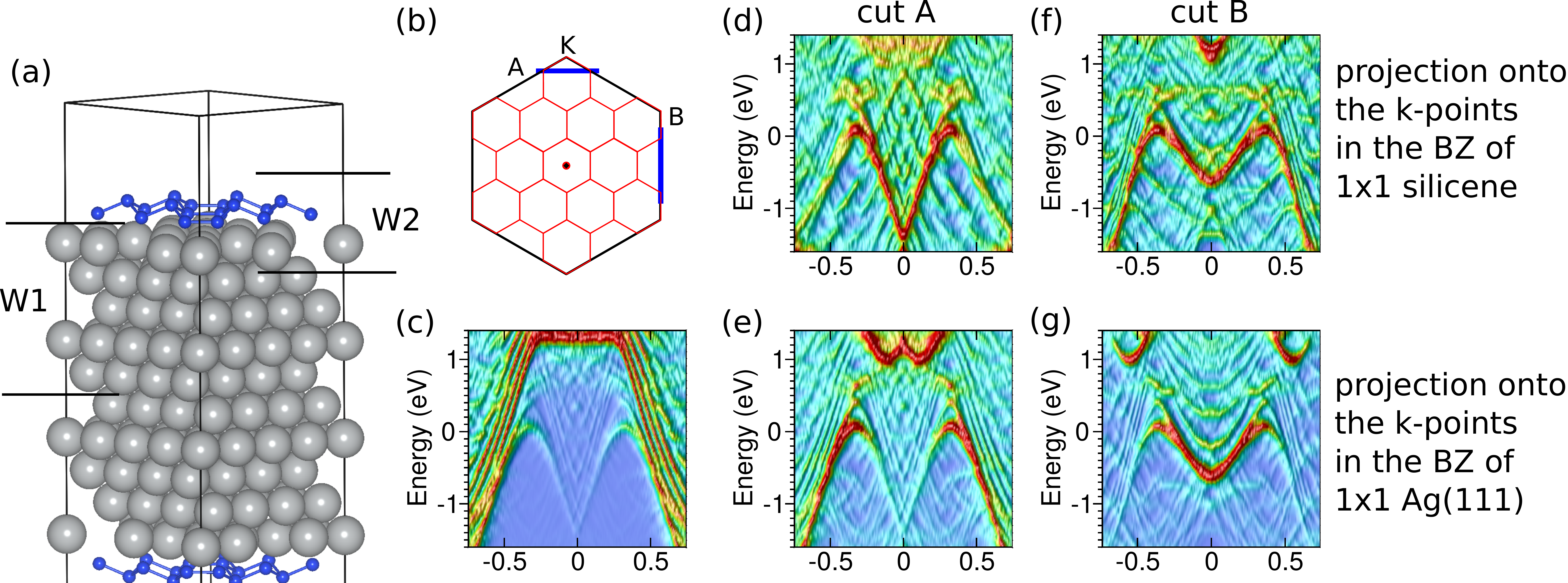}
  \caption{Unfolded band structures for silicene on Ag(111). 
  (a) Geometry of 3$\times$3 silicene on 4$\times$4 Ag(111). 
  W1 and W2 denote the spatial windows, in which the wave functions are chosen for band unfolding.
  W1 contains five layers of Ag(111), while W2 covers only the first Ag layer and the silicene.
  (b) The BZs of the supercell and the 1$\times$1 primitive cell of Ag(111). 
  The blue lines (labeled cuts A and B, respectively) represent the high symmetry lines for band calculations.
  (c) Unfolded band structure along cut A for the substrate by projecting the wave functions in W1 onto the $k$-points
  in the BZ of 1$\times$1 Ag(111).
  (d), (e) Unfolded band structures along cut A for silicene and Ag(111), respectively.
  (f), (g) Unfolded band structures along cut B for silicene and Ag(111), respectively.
  For (d)--(g) wave functions in W2 were chosen for band unfolding.
  The Fermi level is set to zero. 
  }
 \label{fig6}
\end{figure*}

BAs was recently predicted to possess intriguing properties, i.e., a hexagonal structure with a direct gap of about 1.1 eV at the K point 
and a high mobility comparable to that of graphene. \cite{Xie2016Two}
Generating spin-splittings in this system may be useful for designing spintronic devices in future applications.
For such a purpose using magnetic semiconductors to induce spin splittings via the magnetic proximity effect 
has several advantages over doping magnetic atoms, such as preserving the atomic structure of the overlayer and
making the manipulation easily controllable, as demonstrated in the 
successful realization of large spin-exchange splittings and anomalous Hall effect in graphene.\cite{Wei2016,Wang2015}
Here we propose to generate spin-exchange splittings in BAs by making use of the magnetic proximity effect. 
A monolayer of the newly discovered ferromagnetic semiconductor CrI$_3$ was used as the substrate. 
The lattice mismatch of a
2$\times$2 supercell of BAs with CrI$_3$ (experiment: 6.867 \AA) is less than 2\%,
suggesting that vdW epitaxy could grow such heterostructures.  We have examined three configurations for
BAs/CrI$_3$, obtained by shifting BAs along the [-110] direction, Fig.~\ref{fig5}, referred to as S1, S2, S3,
respectively.  DFT calculations find that S3 has the lowest energy, about 9 meV per BAs unit cell lower than
S1.  Layer distances between the BAs and CrI$_3$ monolayers are in the range of 3.80--3.95 \AA.  Unfolded band
structures for BAs were obtained by projecting the supercell wave functions in BAs onto the $k$-points of
1$\times$1 BAs.  The spin splittings are found to be configuration-dependent: negligibly small in the valence
bands for S1 and S3, about 50 meV and 25 meV in the conduction bands for S1 and S3, respectively.  The
spin-splitting for S1 is comparable to the calculated value for graphene/EuO(111) (48 meV for the conduction band),\cite{Hallal2017} 
where the layer distance (2.57 \AA) is much smaller than that for BAs/CrI$_3$. 
Therefore, a large spin-exchange splitting can be
effectively obtained in BAs via magnetic proximity effect in vdW heterostructures. 

\subsection{Interaction induced interface states: silicene on Ag(111)}

Silicene on Ag(111) has received much attention during the past few years.
Unfortunately, the strong interaction between the overlayer and the substrate destroys the Dirac states in silicene.
\cite{Chen2014,guo2013,wang2013Ab,cahangirov_2013,gori_2013,Mahatha_2014}
However, recently an ARPES study observed that there are six pairs of half Dirac cones below the Fermi level
on the edges of the first BZ of Ag(111), other than at the K points of 1$\times$1 silicene.\cite{Feng2016}
This observation led to the claim that Dirac cones exist in this system near the edge of the BZ,
and were attributed to the interaction of the overlayer and the substrate.
To clarify if Dirac states exist as claimed, 
we have performed DFT and
layer $k$-projection calculations to understand 
how the interaction between silicene and Ag(111) affects the electronic bands.
The structural model is the one we used for our previous study, \cite{Chen2014}
for which the simulated STM is in good agreement with the experiments.\cite{Feng2016,Vogt_PRL} 
Figure~\ref{fig6}(b) shows the BZs of the 1$\times$1 and 4$\times$4 Ag(111), and
cuts A and B are the high symmetry lines probed in the ARPES experiments and our $k$-projection calculations.

Since previously\cite{Chen2014} the linear dispersion observed for silicene/Ag(111) was found to originate from the substrate,
we first consider the unfolded band structure along cut A for the substrate, Fig.~\ref{fig6}(c), 
obtained by projecting the supercell wave functions in the spatial window W1
onto the $k$-points in the BZ of 1$\times$1 Ag(111).
There are a few linear-like bands crossing the Fermi level, 
 unlike the bands seen by the ARPES measurement (Fig.\ 3 in Ref.~\onlinecite{Feng2016}).

Since ARPES is surface-sensitive, the experiments of
Ref.~\onlinecite{Feng2016} using a photon energy of 21.218 eV for the ARPES experiments may have detected  
the surface bands of Silicene/Ag(111).
According to Ref.~\onlinecite{arpes-depth} the expected probing depth is $\sim$5 \AA for photons of $\sim$20-22 eV.
Thus, we consider states with weight in W2 (silicene and the first Ag(111) layer) and
$k$-project them onto the BZs of 1$\times$1 of silicene (Figs.~\ref{fig6}(d,f)) and Ag(111)
(Figs.~\ref{fig6}(e,g)).
For silicene an M-shaped band right below the Fermi level can be seen, 
but the V-shape part in the center has higher intensities than the two arms.    
For the substrate, the situation is opposite.
Our results are also consistent with the previous study.\cite{Lian2017Dirac}
We further note that the calculated band structure agrees with the ARPES results (Fig. 3 in Ref.~\onlinecite{Feng2016})
if one superimposes the unfolded band structures for both the silicene (Fig.~\ref{fig6}(d)) and the substrate
(Fig.~\ref{fig6}(e)).
Likewise, our calculations along cut B are also in good agreement with the ARPES experiments.\cite{Feng2016,Mahatha_2014}
However, as shown in Fig.~\ref{fig6}, these are not Dirac states as claimed by the experiment.  
Nonetheless, our results are consistent with the ARPES experiment 
in that these bands are interface states resulting from the interaction between silicene and the substrate. 

\subsection{Bulk-surface decomposition of Dirac states in Bi$_2$Se$_3$}

\begin{figure}
\includegraphics[width=0.95\columnwidth]{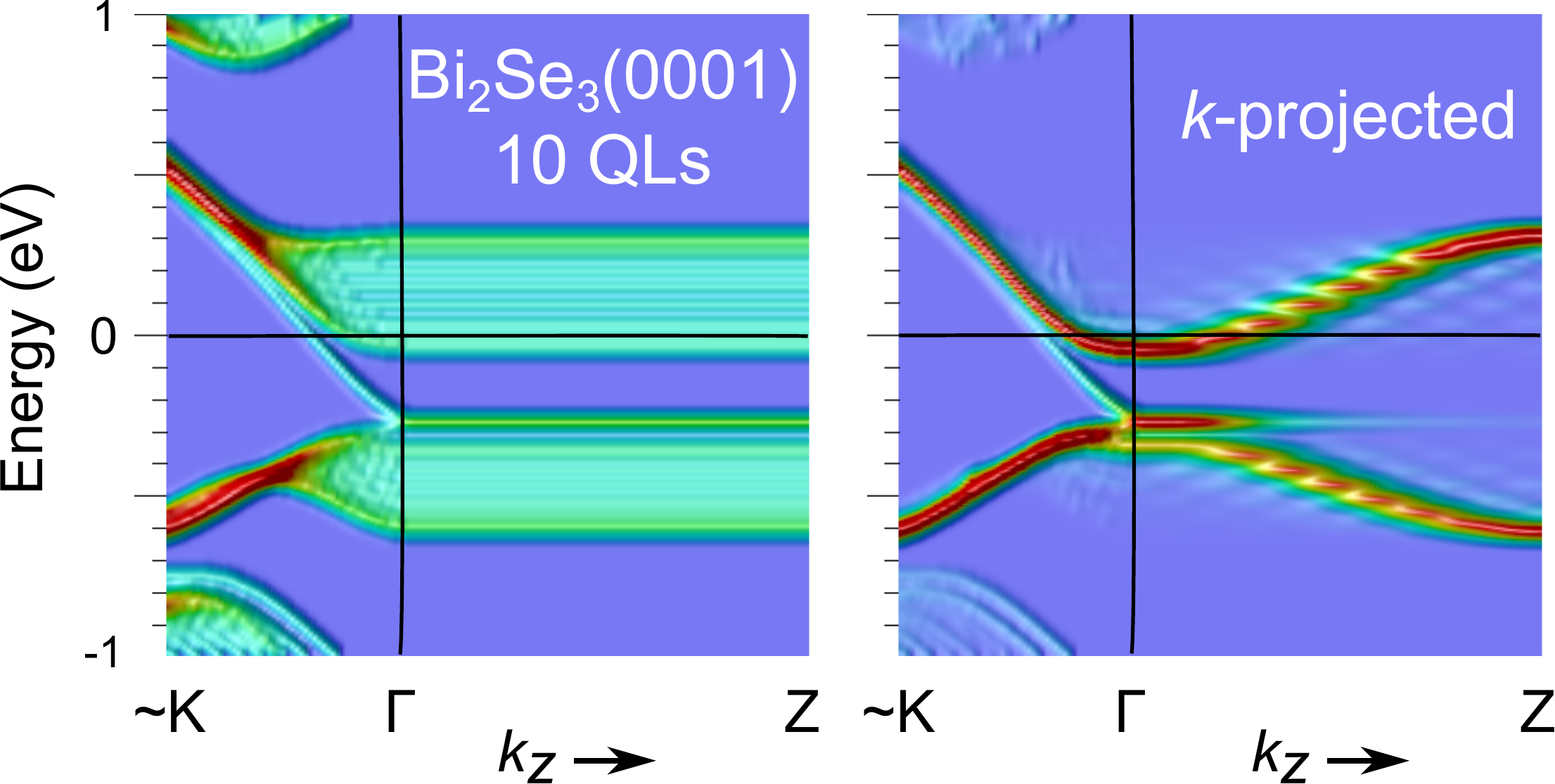}
\caption{Calculated bands, including the topological Dirac state, for a 10 QL film of Bi$_2$Se$_3$(0001) for along K ($\sim$0.2 \AA$^{-1}$) to $\Gamma$
and then perpendicular to Z. The left panel left panel is a layer projection in the top 3 QLs, and the
corresponding bulk $k$-projected band are shown on the right. The Fermi level is set to zero.
}
\label{fig7}
\end{figure}

The topological surface states of Bi$_2$Se$_3$(0001) have been well studied, and calculations (and experiments)
have shown that a minimum of about 5 QLs are necessary for the topological state to form a Dirac cone. For
thinner films, there are still surface states, but are of the ``normal'' variety that
can be understood as splitting off
from the bulk bands. Since there is an evolution of the Dirac with thickness, there should be a connection
between the models of standard and topological surface states. Here briefly analyze this connection.
We consider 10 QLs of Bi$_2$Se$_3$. The calculated surface $k_\|$ bands, Fig.~\ref{fig7} show the Dirac state, and a continuum
of valence and conduction bands that result from the projection of the bulk bands. For the surface
(or in a repeated slab) calculations, the translational symmetry perpendicular to the surface is broken, but
the wave functions can still be labeled by $k_z$; in this case, if the calculations are converged and there are
no artificial interactions between images, then there should be no variation in the calculations with $k_z$. As
seen in the left panel, this is indeed the case. $k$-projecting these bands to the bulk cell (1 QL), $k_z$=0
for K-$\Gamma$ and $k_\|$=0 for $\Gamma$-Z, yields well-define bulk states, albeit the minigaps in the valence
and conduction band along $\Gamma$-Z reflecting the finite number of layers. The results show that the Dirac
state is built up mainly of states with $k_z$ around $\Gamma$ split off from the valence band. This result
shows that the standard arguments for normal surface state formation also hold for the topological surface
states; for the topological states, the band inversion affects the character of the valence states out to about
0.2 of $\Gamma$-Z, which in turn then contribute to the Dirac state; for fewer than 5 QLs, the surface state is
seen to split off the conduction band, in  which the minigaps are also larger because of the smaller number of
layers.

\section{CONCLUSIONS}
In summary, we have presented a technique for unfolding electronic bands of materials, including
an efficient scheme using FFTs to calculate the local $k$-projected bands.
This method allows us to effectively study the effects of interfaces by 
examining the spatial characteristics of the band structure,
which is useful for understanding ARPES and STM/STS experiments.
We applied the method to interfaces systems of gr-2L/SiC(0001), BAs/CrI$_3$, silicene/Ag(111), and
Bi$_2$Se$_3$/vaccuum. 
Our results revealed that the interactions of gr-2L and the two surfaces of SiC(0001) behave differently:
The Si-face has minor effects on the band structure of the gr-2L, with a gap at the K point of only about 10
meV; on the C-face, however, a gap of about 130 meV at K is induced and gr-2L is $n$-doped, caused
by a strong electric dipole at the interface caused by a charge polarization.
For the vdW heterostructure BAs/CrI$_3$, we showed that the magnetic proximity effect can cause spin splittings of
up to 50 meV  in BAs that depend on the lateral registry of the two layers.
For silicene/Ag(111), our results are consistent with recent ARPES experiments that find interface states whose
dispersions on the edge of the first Brillouin zone of Ag(111) appear to be half Dirac cones,
but demonstrate that they are not Dirac states. Finally, we have shown that the $k$-projection can provide
insight into the bulk origin of surface states, including the topological Dirac states.

\begin{acknowledgments}
This work was supported by the National Natural Science Foundation of China (Grants No. 11774084) ,
and the U.S. National Science Foundation, EFMA-1741673.
\end {acknowledgments}

\bibliography{references}
\bibliographystyle{apsrev4-1}
\end{document}